\documentclass[aps,prb,showpacs,amssymb,preprint,floatfix]{revtex4}
\usepackage{amssymb}
\usepackage{graphicx}
\usepackage{color}
\usepackage[normalem]{ulem}
\begin{document}

\title{Giant crystal-electric-field effect and complex magnetic behavior in single-crystalline CeRh$_3$Si$_2$}

\author{A.~P.~Pikul, D. Kaczorowski, Z. Gajek, J. St{\c{e}}pie{\'{n}}--Damm}
\affiliation{Institute of Low Temperature and Structure Research, Polish Academy of Sciences, P Nr
1410, 50--950 Wroc{\l}aw, Poland}

\author{A.~\'Slebarski}
\affiliation{Institute of Physics, University of Silesia, 40--007 Katowice, Poland}

\author{M.~Werwi\'nski, A.~Szajek}
\affiliation{Institute of Molecular Physics, Polish Academy of Sciences, 60--179 Pozna{\'n}, Poland}
\date{\today}

\begin{abstract}
Single-crystalline CeRh$_3$Si$_2$ was investigated by means of x-ray diffraction, magnetic
susceptibility, magnetization, electrical resistivity, and specific heat measurements carried out
in wide temperature and magnetic field ranges. Moreover, the electronic structure of the compound
was studied at room temperature by cerium core-level x-ray photoemission spectroscopy (XPS). The
physical properties were analyzed in terms of crystalline electric field and compared with results
of ab-initio band structure calculations performed within the density functional theory approach.
The compound was found to crystallize in the orthorhombic unit cell of the ErRh$_3$Si$_2$ type
(space group Imma -- No.74, Pearson symbol: oI24) with the lattice parameters: $a =
7.1330(14)$~{\AA}, $b = 9.7340(19)$~{\AA}, and $c = 5.6040(11)$~{\AA}. Analysis of the magnetic and
XPS data revealed the presence of well localized magnetic moments of trivalent cerium ions. All
physical properties were found to be highly anisotropic over the whole temperature range studied,
and influenced by exceptionally strong crystalline electric field with the overall splitting of the
4$f^1$ ground multiplet exceeding 5700 K. Antiferromagnetic order of the cerium magnetic moments at
$T_{\rm N}$~=~4.70(1)~K and their subsequent spin rearrangement at $T_{\rm t}$~=~4.48(1)~K manifest
themselves as distinct anomalies in the temperature characteristics of all investigated physical
properties and exhibit complex evolution in an external magnetic field. A tentative magnetic $B\! -
\!T$ phase diagram, constructed for $B$ parallel to the $b$-axis being the easy magnetization
direction, shows very complex magnetic behavior of CeRh$_3$Si$_2$, similar to that recently
reported for an isostructural compound CeIr$_3$Si$_2$. The electronic band structure calculations
corroborated the antiferromagnetic ordering of the cerium magnetic moments and well reproduced the
experimental XPS valence band spectrum.

\end{abstract}

\pacs{75.10.Dg (crystal-field theory and spin Hamiltonians), 75.30.Gw (magnetic anisotropy),
75.30.Kz (magnetic transitions), 75.50.Ee (antiferromagnetics), 71.20.Lp (electronic structure of
bulk materials)} \maketitle


\section{\label{introduction_section} Introduction}

The phase diagram Ce-Rh-Si comprises twenty five ternary phases with well defined crystal
structures [\onlinecite{lipatov}]. However, physical properties have been reported for only a few
of them, namely: CeRhSi$_2$ and Ce$_2$Rh$_3$Si$_5$ (both described in the literature as
intermediate valence systems [\onlinecite{adroja,godart,kaczorowski3}]), Ce$_3$Rh$_3$Si$_2$
(exhibiting complex long-range magnetic ordering [\onlinecite{kaczorowski1}]), CeRh$_2$Si$_2$ and
CeRhSi$_3$ (antiferromagnetic heavy--fermion systems [\onlinecite{settai,muro}]). The two latter
phases are known as pressure-induced superconductors
[\onlinecite{araki,movshovich,kimura1,kimura2}]. While in CeRh$_2$Si$_2$ the superconductivity
seems to replace the antiferromagnetism [\onlinecite{araki,movshovich}], in CeRhSi$_3$ it develops
deeply inside the ordered state [\onlinecite{kimura1,kimura2}]. Most interesting, the
superconductivity in non-centrosymmetric CeRhSi$_3$ involves spin-fluctuations-assisted triplet
pairing [\onlinecite{kimura2}]. In the context of recent findings for the Ce--Rh--Si ternaries,
comprehensive physical characterization of other (hitherto unknown or hardly studied) phases from
the system seems quite desirable. To the best of our knowledge existence of CeRh$_3$Si$_2$ has for
the first time been reported by Morozkin et al. [\onlinecite{morozkin}], yet without any
crystallographic or physical data. Then, CeRh$_3$Si$_2$ has been found as an impurity phase in
polycrystalline CeRh$_2$Si$_2$ [\onlinecite{graf}]. In the latter study it has been established
that the compound crystallizes with an orthorhombic crystal structure and exhibits an
antiferromagnetic phase transition at $T_{\rm N}$ = 5 K [\onlinecite{graf}].

Here we report the results of our investigation of the crystal structure and the basic physical
properties of CeRh$_3$Si$_2$ performed on single-crystalline specimens in wide temperature and
magnetic field ranges. The experimental characterization is accompanied by the results of ab-initio
calculations of the electronic band structure. Short accounts on the work presented in this paper
were given in conference communications [\onlinecite{kaczorowski2, pikul}].


\section{\label{experimental_details_section} Experimental and computational details}

A single crystal of CeRh$_3$Si$_2$ was grown by the Czochralski pulling method employing a
tetra-arc furnace under protective ultra-pure (4.8N, BOC Gases) argon atmosphere, which was
additionally titanium-gettered during the whole growing process. The starting polycrystalline melt
was prepared from pieces of Ce (3N, Ames Laboratory), Rh ingot (3N, Chempur) and Si chips (6N,
Chempur). The pulling rate was 10~mm/h, and the copper heart rotation speed was 3~rpm. The final
ingot of CeRh$_3$Si$_2$ was about 4~mm in diameter and 40~mm in length. From this
single-crystalline rod a fragment of about 20~mm in length was cut for experimental studies. The
crystal was wrapped in Ta-foil, sealed in an evacuated silica tube, and annealed at 900$^{\circ}$C
for 2 weeks. The quality of the product was verified by means of x-ray powder diffraction
measurements performed on a pulverized piece of the crystal (Stoe diffractometer with Cu
K${\alpha}$ radiation) and microprobe analysis (Phillips 515 scanning electron microscope equipped
with an EDAX PV 9800 spectrometer). The entire X-ray diffractogram was indexed with an orthorhombic
unit cell, and the sample was found to be single phase. Additionally, a polycrystalline sample of
an isostructural LaRh$_3$Si$_2$ compound, being a non-magnetic counterpart of CeRh$_3$Si$_2$, was
prepared by conventional arc melting stoichiometric amounts of the constituents. X-ray powder
diffraction experiments proved that the La-based phase is indeed isostructural to CeRh$_3$Si$_2$.

The crystal structure of CeRh$_3$Si$_2$ was also examined on a small single-crystalline fragment
using a four-circle diffractometer equipped with a CCD camera (Kuma Diffraction KM-4 with
graphite-monochromatized Mo K${\alpha}$ radiation). Crystal structure refinement was performed
using full-matrix least-squares on $F^2$ as a refinement method and applying the SHELXL-97 program
[\onlinecite{shelxl}]. The magnetic properties were studied at temperatures ranging from 1.7~K up
to 300~K in applied magnetic fields up to 5 T using a Quantum Design MPMS-5 superconducting quantum
interference device (SQUID) magnetometer. The electrical resistivity was measured in the
temperature range 1.5 -- 300 K employing standard DC four-point technique. Electrical contacts to
bar-shaped specimens were made by spot welding. The heat capacity was measured in the temperature
interval 350 mK -- 300 K in applied magnetic fields up to 9 T using a Quantum Design PPMS platform.
For these experiments the thermal relaxation method [\onlinecite{hwang}] was applied with 2\% heat
pulses and Apiezon N vacuum grease as a sample mounting medium. The x-ray photoelectron
spectroscopy (XPS) measurements were carried out at room temperature with monochromatized Al
K$\alpha$ radiation using a PHI-5700 ESCA spectrometer. The sample of CeRh$_3$Si$_2$ was scraped
with a diamond file under high vacuum immediately before recording a spectrum.

The electronic structure calculations were performed with a density functional theory (DFT)
[\onlinecite{as01}] using the full-potential linearized augmented plane wave (FP LAPW) method
implemented in the latest version (Wien2k) of the original WIEN code [\onlinecite{as02}]. In the
Wien2k code calculations the scalar relativistic approach was implemented with the spin-orbit
interactions taken into account by using the second variational method [\onlinecite{as03}]. The
exchange correlation potential in the local (spin) density approximation (LSDA) was assumed in the
form proposed by Perdew, Burke, and Ernzerhof [\onlinecite{as04}] using the generalized gradient
approximation. Furthermore, to improve the description of the strongly correlated 5f electrons, the
on-site Coulomb energy $U$ correction and exchange $J$ parameters were introduced within the LSDA+U
approach [\onlinecite{as05a,as05b}]. In our case we used $U_{\rm eff}$~=~$U-J$, setting $J=0$.
Another approach applied by us to reduce the discrepancy between the DFT magnetic moments and
experimental ones was taking into account the so-called orbital polarization (OP) term, as proposed
by Brooks [\onlinecite{as06}] and Eriksson et al. [\onlinecite{as07a,as07b}]. The number of
$k$-points was 2000 in the Brillouin zone (BZ), which corresponds to 240 points in an irreducible
wedge of the BZ for all methods of calculations applied in this paper. For the BZ integration, a
tetrahedron method was used [\onlinecite{as08}]. The self-consistency criterion was equal 10$^{-6}$
Ry for the total energy. The calculations were performed for lattice constants and atomic positions
in the unit cell described below in Sec.~\ref{structure_section}. The theoretical x-ray
photoemission spectra were obtained from the calculated densities of electronic states (DOS)
convoluted by a Gaussian with a half-width $\delta$ equal to 0.3~eV and scaled using the proper
photoelectronic cross sections for partial states [\onlinecite{as09}].


\section{\label{results_discussion_section} Results and discussion}
\subsection{\label{structure_section} Crystal structure}

\begin{figure}
\includegraphics[width=8.6cm]{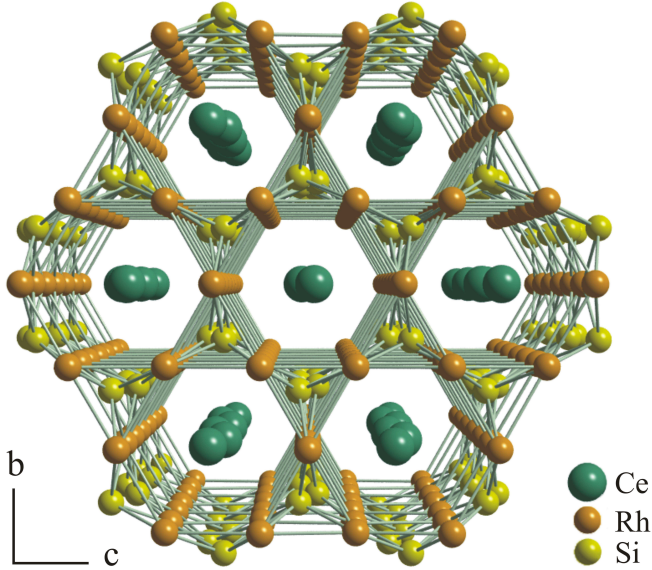}
\caption{\label{structure_figure} (Color online) Crystal structure of CeRh$_3$Si$_2$.}
\end{figure}

\begin{table}
\caption{\label{structure_table} Crystallographic and structure refinement data for
single-crystalline CeRh$_3$Si$_2$.}
\begin{ruledtabular}
\begin{tabular}{l l}
Empirical formula & Ce$_{\rm 0.50}$ Rh$_{1.50}$ Si \\
Formula weight & 252.51 g/mol \\ Temperature & 293(2) K \\
Wavelength & 0.71073 \AA \\
Crystal system, space group & orthorhombic, Imma \\
Unit cell dimensions & $a$ = 7.1330(14) {\AA}, $\alpha = 90^{\circ}$ \\
& $b$ = 9.7340(19) {\AA}, $\beta = 90^{\circ}$ \\
& $c$ = 5.6040(11) {\AA}, $\gamma = 90^{\circ}$ \\
Unit cell volume & 389.10(13) {\AA}$^3$ \\ $Z$ & 4 \\
Calculated density & 8.621 Mg/m$^3$ \\
Absorption coefficient & 24.354 mm$^{-1}$ \\
$F (000)$ & 884 \\
$\theta$ range for data collection & $4.19^{\circ} \div  45.28^{\circ}$ \\
Limiting indices & $-13\leq h \leq 9$ \\
& $-12\leq k \leq 19$ \\
& $-10 \leq l \leq 10$ \\
Reflections collected / unique  & 4366 / 840 $[R(\rm{int}) = 0.0465]$ \\
Completeness to $\theta$ = 45.28 & 93.8 \% \\
Refinement method & Full-matrix least-squares on $F^2$ \\
Data / restraints / parameters & 840 / 0 / 22 \\
Goodness-of-fit on $F^2$ & 1.010 \\
Final $R$ indices $[I>2 \sigma (I)]$ & $R_1 = 0.0287$, $wR_2 = 0.0509$ \\
$R$ indices (all data) & $R_1 = 0.0432$, $wR_2 = 0.0524$ \\
Extinction coefficient & 0.0356(8) \\
Largest diff. peak and hole & 2.707 e{\AA}$^{-3}$ and $-3.148$ e{\AA}$^{-3}$
\end{tabular}
\end{ruledtabular}
\end{table}

\begin{table}
\caption{\label{coordinates_table} Atomic coordinates and equivalent isotropic thermal displacement
parameters for CeRh$_3$Si$_2$. $U_{\rm eq}$ was defined as one third of the trace of the
orthogonalized $U_{ij}$ tensor.}
\begin{ruledtabular}
\begin{tabular}{l c c c c} Atom & $x$ $[\times 10^4]$ & $y$ $[\times 10^4]$ & $z$
$[\times 10^4]$ & $U_{\rm eq}$ $[\times 10^3 \text{\AA}^2]$\\
\hline Ce & 0 & 7500 & 2160(1) & 8(1) \\
Rh(1) & 2172(1) & 5000 & 0 & 6(1) \\
Rh(2) & $-2500$ & 7500 & $-2500$ & 8(1) \\
Si & 0 & 9215(2) & 6996(2) & 7(1) \\
\end{tabular}
\end{ruledtabular}
\end{table}

\begin{table}
\caption{\label{displacement_table} Anisotropic thermal displacement parameters ($\times 10^3
\text{\AA}^2$) for CeRh$_3$Si$_2$. The anisotropic displacement factor exponent takes the form: $-2
\pi^2 [h^2 a^2 U_{\rm 11} + ... + 2 h k a b U_{\rm 12}]$.}
\begin{ruledtabular}
\begin{tabular}{l c c c c c c} Atom & $U_{\rm 11}$ & $U_{\rm 22}$ & $U_{\rm 33}$ &
$U_{\rm 23}$ & $U_{\rm 13}$ & $U_{\rm 12}$ \\
\hline Ce & 8(1) & 7(1) & 8(1) & 0 & 0 & 0 \\
Rh(1) & 7(1) & 6(1) & 5(1) & 1(1) & 0 & 0 \\
Rh(2) & 13(1) & 4(1) & 7(1) & 0 & $-2(1)$ & 0 \\
Si & 6(1) & 9(1) & 7(1) & $-1(1)$ & 0 & 0
\end{tabular}
\end{ruledtabular}
\end{table}

The crystal structure refinement (see Tab.~\ref{structure_table}) performed for single-crystalline
CeRh$_3$Si$_2$ confirmed that the compound crystallizes in the orthorhombic symmetry with the space
group Imma (No.74, Pearson symbol oI24) and that it is isostructural with ErRh$_3$Si$_2$
[\onlinecite{cenzual}]. Refined lattice parameters are: $a$ = 7.1330(14) {\AA}, $b$ = 9.7340(19)
{\AA}, $c$ = 5.6040(11) {\AA}, and the atomic coordinates are listed in
Tab.~\ref{coordinates_table}. Figure~\ref{structure_figure} presents the structure of
CeRh$_3$Si$_2$ viewed along the $a$-axis.

The ErRh$_3$Si$_2$-type crystal structure was described in detail in Ref.~\onlinecite{cenzual}. It
represents a deformed superstructure of the hexagonal CaCu$_5$-type structure with $a = 2c_h$, $b =
\sqrt{3}a_h$, and $c = a_h$, where $a_h$ and $c_h$ are the lattice parameters of the parent
hexagonal unit cell. The characteristic features of this atom arrangement are a honeycomb-like
network of the Rh and Si atoms within the $bc$ plane and zigzag chains of the Ce atoms along the
$a$-axis (see Fig.~\ref{structure_figure}).

\subsection{\label{magnetic_properties_section} Magnetic properties}

\begin{figure}
\includegraphics[width=8.6cm]{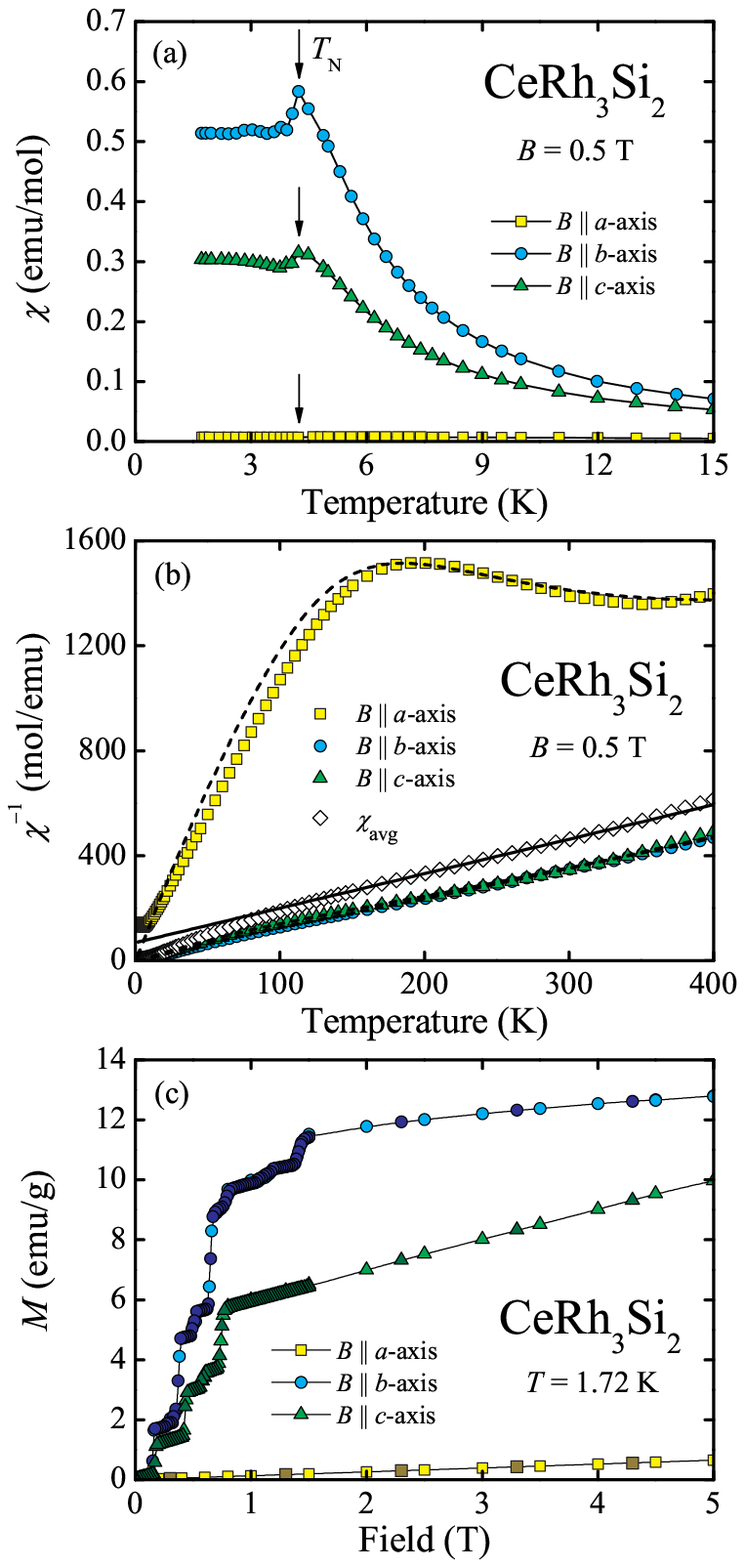}
\caption{\label{susceptibility_figure} (Color online)
(a) Low-temperature magnetic susceptibility of single-crystalline CeRh$_3$Si$_2$ measured in
external magnetic field $B$ applied parallel to the principal crystallographic axes. The arrows
mark the N{\'{e}}el temperature; the solid lines serve as a guide for the eye. (b) Temperature
variations of the inverse magnetic susceptibility of the compound; $\chi_{\rm avg}^{-1}$
represents the inverse average magnetic susceptibility. The solid line is a fit of the Curie--Weiss law
to the experimental data; the dashed curves represent the
fits described in Sec.~\ref{CEF_calc}. (c) Magnetization as a function of applied magnetic field
measured for single-crystalline CeRh$_3$Si$_2$ in zero-field (bright symbols) and field
cooling (dark symbols) regimes, along the principal crystallographic axes. The solid lines serve
as guides for the eye.}
\end{figure}

\begin{figure}
\includegraphics[width=8.6cm]{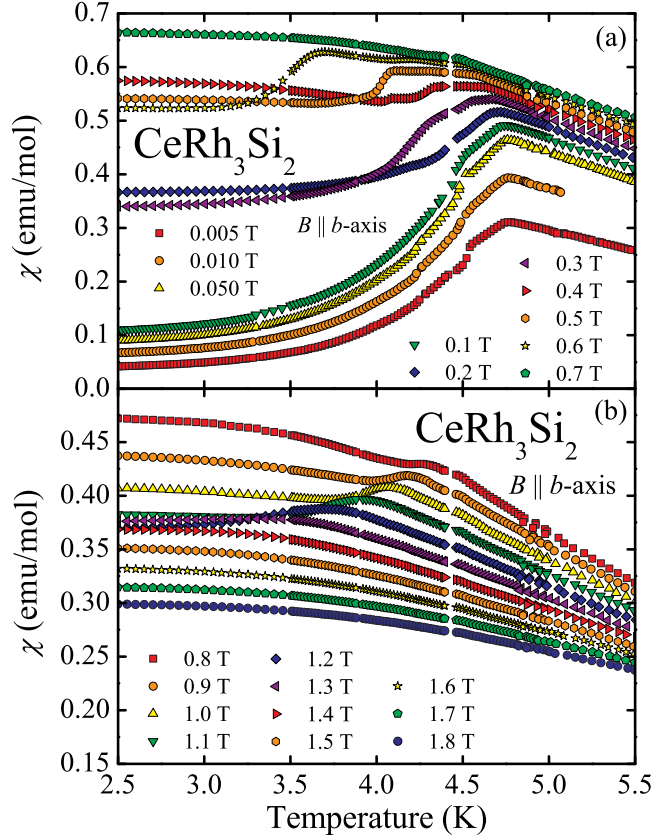}
\caption{\label{susceptibility_low_temperatures_figure} (Color online) Temperature
dependencies of the magnetic susceptibility of single-crystalline CeRh$_3$Si$_2$ measured in
different applied magnetic fields $B$ parallel to the $b$-axis. For the sake of
clarity the curves are shifted up.}
\end{figure}

\begin{figure}
\includegraphics[width=8.6cm]{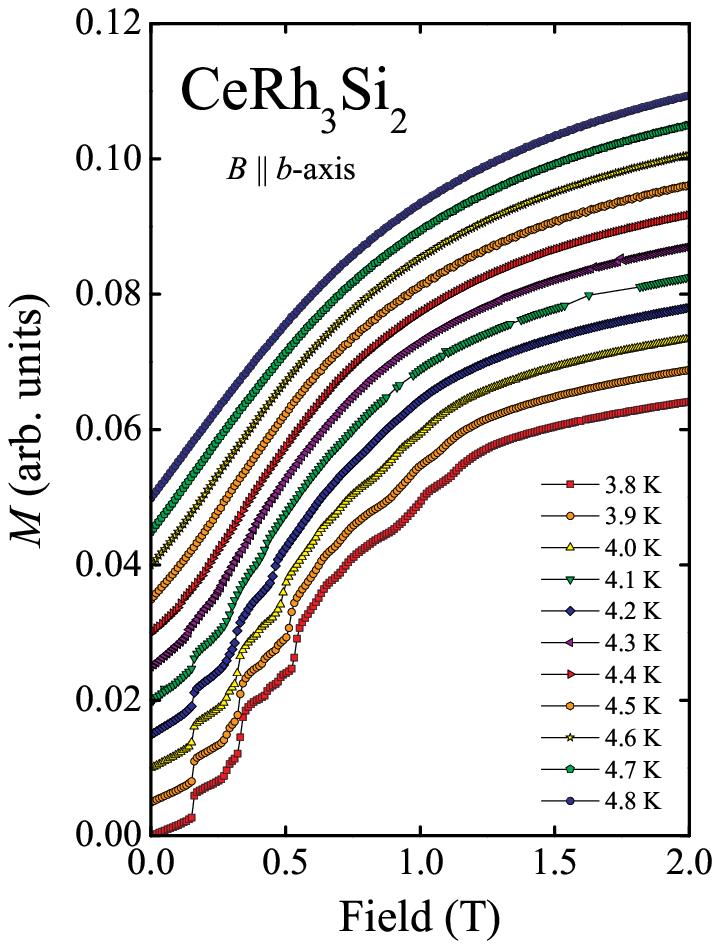}
\caption{\label{magnetization_low_temperatures_figure} (Color online) Isothermal
magnetization of single-crystalline CeRh$_3$Si$_2$ measured with the magnetic field applied
along the $b$-axis at several different temperatures from the ordered region. The curves are
shifted up for the sake of clarity.}
\end{figure}

Figures~\ref{susceptibility_figure}(a,b) present the magnetic susceptibility $\chi$ of
CeRh$_3$Si$_2$ measured as a function of temperature $T$ in a magnetic field $B$ of 0.5~T applied
parallel to the main crystallographic directions. Sharp susceptibility maxima at about 4.5~K, best
visible on the $\chi_b (T)$ curve measured along the $b$-axis [see
Fig.~\ref{susceptibility_figure}(a)], manifest the onset of an antiferromagnetically ordered state.
Below 4.5~K the susceptibility $\chi_b$ does not decrease significantly, as expected for simple
antiferromagnets, but saturates at low temperatures, hence suggesting more complex magnetic
behavior. The anomaly at 4.5~K is only slightly weaker for the $c$-axis, and the behavior of
$\chi_c$ in the ordered region is similar to that of $\chi_b$. On the contrary, the $a$-axis
susceptibility is featureless and very small, thus indicating that the magnetic moments are
confined to the $bc$ plane of the unit cell. A distinct difference of $\chi_a (T)$ as compared to
the other two components is observed also in the paramagnetic state [see
Fig.~\ref{susceptibility_figure}(b)]. At room temperature, the ratio $\chi_b / \chi_a$ is about 4,
and it increases with decreasing temperature down to 200~K, where it is larger than 6. The
$\chi^{-1}_a (T)$ variation exhibits strongly curvilinear shape in the whole $T$-range studied. In
contrast, the $\chi^{-1}_b (T)$ and $\chi^{-1}_c (T)$ curves are almost identical and exhibit
nearly linear behavior characteristic of systems with well localized magnetic moments. The overall
behavior of the magnetic susceptibility of CeRh$_3$Si$_2$ implies huge magnetocrystalline
anisotropy due to strong crystal field interactions (cf. Sec.~\ref{CEF_calc}).

As can be seen in Fig.~\ref{susceptibility_figure}(b), the inverse average "polycrystalline"
magnetic susceptibility $\chi_{\rm avg}^{-1}$ [defined as $\chi_{\rm avg} = \frac{1}{3} (\chi_a +
\chi_b + \chi_c)$] follows above about 150~K the Curie--Weiss law with the effective magnetic
moment $\mu_{\rm eff} = 2.47$~$\mu_{\rm B}$ and the paramagnetic Curie temperature $\theta_{\rm CW}
= -52$~K [see the straight solid line in Fig.~\ref{susceptibility_figure}(b)]. The experimental
value of $\mu_{\rm eff}$ is close to that calculated for a free Ce$^{3+}$ ion
($g\sqrt{j(j+1)}=2.54$), and thus indicates the presence of well localized magnetic moments. The
negative value of $\theta_{\rm CW}$ hints at antiferromagnetic character of the magnetic exchange
interactions, in line with the AFM ordering observed below 4.5~K.

Figure~\ref{susceptibility_figure}(c) shows the magnetization of CeRh$_3$Si$_2$ measured at 1.72~K
as a function of the magnetic field strength. Here the complex magnetic behavior of the compound
manifests itself as cascades of metamagnetic-like transitions in $M (B)$ measured along the $b$-
and $c$-axes. For the former direction of the magnetic field one observes some saturation above
about 1.5~T at a value $\mu_{\rm s}$~=~12.8~emu/g corresponding to about 1.16~$\mu_{\rm B}$. In the
case of $B \parallel c$-axis, the magnetization increases linearly with increasing $B$ reaching in
5 T a value of about 0.9~$\mu_{\rm B}$. In turn, for $B \parallel a$-axis the magnetization is a
linear function of the applied field in the entire range studied, and the value of $M$ measured at
5 T is nearly 20 times smaller than that observed for $B \parallel b$-axis. The overall behavior of
$M (B)$ is reminiscent of that reported for CeSb [\onlinecite{rossat-mignod1,rossat-mignod2}],
TbNi$_2$Si$_2$ [\onlinecite{shigeoka}], and TbNi$_2$Ge$_2$ [\onlinecite{budko}], which are commonly
called "devil's staircase" systems (for short review and discussion see Sec.~\ref{summary}).

In order to shed more light on the nontypical temperature variation of the magnetic susceptibility
below 4.5~K [Fig.~\ref{susceptibility_figure}(a)] we performed a series of additional measurements
of $\chi_{\rm b}(T)$, with increased density of the data points. As can be inferred from
Fig.~\ref{susceptibility_low_temperatures_figure}, in the lowest magnetic field studied (i.e.
$B$~=~0.005~T) the magnetic order in CeRh$_3$Si$_2$ sets in already at $T_{\rm N}$~=~4.7~K. The
N\'{e}el temperature decreases with increasing field, as expected for antiferromagnets. In a field
of 0.5~T, $T_{\rm N}$ is reduced to about 4.5~K, marked by an arrow in
Fig.~\ref{susceptibility_figure}(a). Moreover, the high resolution measurements revealed the
presence of some spin rearrangement, which takes place just below $T_{\rm N}$ and therefore is
hardly visible in Fig.~\ref{susceptibility_figure}(a). In 0.005~T, this spontaneous change of the
magnetic structure manifests itself as a tiny anomaly in the $\chi (T)$ curve at $T_{\rm
t}$~=~4.5~K, which develops with increasing field into a clear kink moving down on the temperature
scale. The latter observation suggests that the antiferromagnetic character of the magnetic order
is retained below $T_{\rm t}$. The plateau visible in 0.5~T below 4~K
[Fig.~\ref{susceptibility_figure}(a)] apparently results from superposition of low-temperature
slopes of the anomalies at $T_{\rm N}$ and $T_{\rm t}$, and another, field-induced phase
transition. It is well visible in Fig.~\ref{susceptibility_low_temperatures_figure} in the field
range 0.7~T~$\leqslant B \leqslant$~0.9~T as a broad ferromagnetic-like hump dominating the
susceptibility below $T_{\rm t}$. Unfortunately, complex shape of the $\chi (T)$ curves does not
allow us to determine the exact temperature at which the latter contribution sets in.

Figure~\ref{magnetization_low_temperatures_figure} presents a series of dense magnetization curves
measured along the easy $b$-axis. For the sake of straightforward comparison with the
afore-discussed susceptibility data, these experiments were focused on magnetic fields lower than
2~T and temperatures close to $T_{\rm N,t}$. As seen, the "devilish" field dependence of the
magnetization, observed at 1.7~K [Fig.~\ref{susceptibility_figure}(c)], is hardly visible at
temperatures higher than 4.4~K. However, upon decreasing $T$ a sequence of distinct steps in the $M
(B)$ curve emerges from the smooth curve. Three of these steps, located at 3.8~K at about 0.15,
0.32 and 0.53~T, dominate the magnetization, yet several less visible anomalies can also be found
on the $M (B)$ curve. The experimental data displayed on
Figs.~\ref{susceptibility_low_temperatures_figure} and \ref{magnetization_low_temperatures_figure}
were used for constructing a tentative phase diagram described in Sec.~\ref{summary}.

\subsection{\label{electrical_resistivity_section} Electrical resistivity}

\begin{figure}
\includegraphics[width=8.6cm]{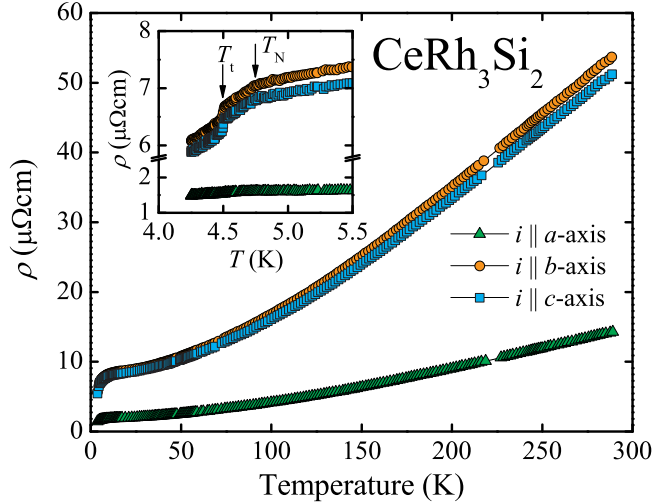}
\caption{\label{electrical_resistivity_figure} (Color online) Electrical resistivity of
single-crystalline CeRh$_3$Si$_2$ measured as a function of temperature with electrical
current $i$ flowing along the principal crystallographic directions; solid lines serve as a guide
for the eye. The inset: low temperature region; the arrows mark magnetic transition
temperatures.}
\end{figure}

Figure~\ref{electrical_resistivity_figure} presents the electrical resistivity of
single-crystalline CeRh$_3$Si$_2$ as a function of temperature, measured with the current $i$
flowing along the main crystallographic directions. Clearly, the compound exhibits good metallic
conductivity in the entire temperature range studied. The antiferromagnetic phase transition
manifests itself as a drop in $\rho (T)$ at about 4.7~K, being equal to the N{\'{e}}el temperature
deduced from the magnetic measurements performed in the weakest magnetic fields (cf.
Fig.~\ref{susceptibility_low_temperatures_figure}). Closer look at the low-temperature range [see
the inset to Fig.~\ref{electrical_resistivity_figure}] unambiguously shows the presence of another
kink on the $\rho (T)$ curve that occurs at about 4.5~K. The latter finding is in-line with the
scenario of two subsequent magnetic phase transitions discussed in
Sec.~\ref{magnetic_properties_section}.

As can be inferred from Fig.~\ref{electrical_resistivity_figure}, the resistivity measured along
the $a$-axis is much smaller than $\rho$ taken along the $b$- and $c$-axes. At 300~K, the ratio
$\rho_b/\rho_a \approx \rho_c/\rho_a \cong$ 4. Moreover, the anomalies at $T_{\rm N}$ and $T_{\rm
t}$ are hardly visible on the $\rho_a (T)$ curve. This behavior is fully consistent with the
established for CeRh$_3$Si$_2$ very large magnetocrystalline anisotropy, characterized by the
magnetic moments confined to the $bc$ plane.

Our several attempts to measure the resistivity of polycrystalline LaRh$_3$Si$_2$ failed because of
multitude of microcracks present in the specimens. For this reason it was not possible to extract
the magnetic contribution to the electrical resistivity of CeRh$_3$Si$_2$ employing a commonly used
method of subtracting the resistivity of its non-magnetic counterpart.

\subsection{\label{specific_heat_section} Specific heat}

\begin{figure}
\includegraphics[width=8.6cm]{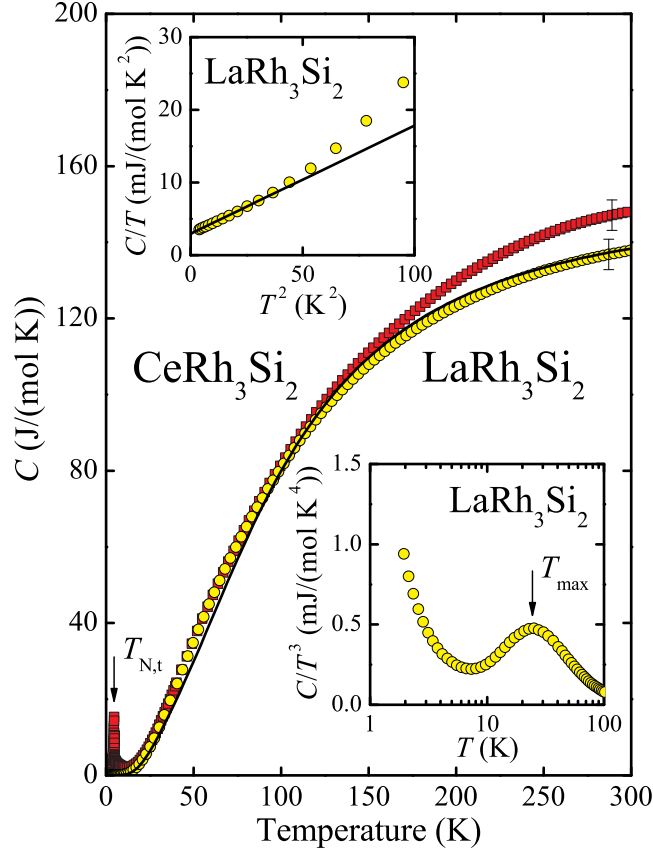}
\caption{\label{specific_heat_figure} (Color online) (a) Specific heat of
single-crystalline CeRh$_3$Si$_2$ (squares) and polycrystalline LaRh$_3$Si$_2$ (circles) as
a function of temperature; the solid line is a fit of Eq.~(\ref{debye_einstein_gamma_equation})
to the experimental data, the arrow marks the N{\'{e}}el temperature. The upper inset shows
$C/T$ vs. $T^2$ measured for the La-based compound; the straight solid line is a fit of the Debye
$T^3$-law [Eq.~(\ref{debye_low_temperatures_equation})]. The lower inset displays the
specific heat of LaRh$_3$Si$_2$ divided by $T^3$ as a function of $T$; the arrow marks the position
of a maximum (for its interpretation see the text).}
\end{figure}

\begin{figure}
\includegraphics[width=8.6cm]{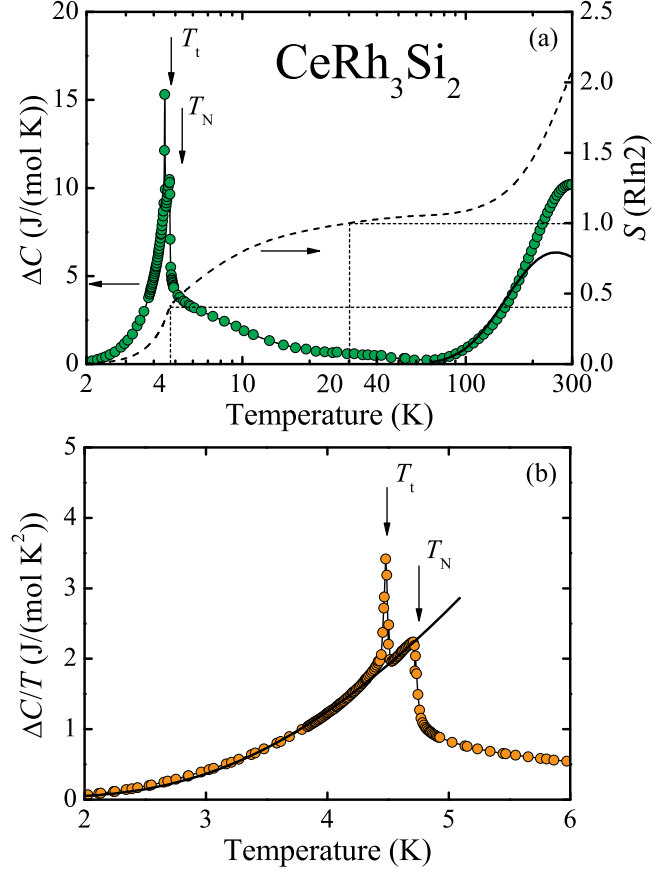}
\caption{\label{specific_heat_magnetic_figure} (Color online) (a) Logarithmic temperature variation of
the magnetic contribution to the specific heat of CeRh$_3$Si$_2$ (left axis); the
thin solid curve serves as a guide for the eye, the arrows mark the magnetic transition
temperatures, and the thick solid line represents the Schottky contribution calculated for the CEF level scheme  from
Tab.~\ref{cef_parameters_table}. The dashed curve is the magnetic entropy $S$ of the compound
(right axis); the drop lines mark the values of $S(T_{\rm N})$ and $T(S=R\ln{2})$. (b) $\Delta C /T$
at low temperatures; the solid curve is a fit of Eq.~(\ref{magnons_equation}) to the experimental
data in the magnetically ordered region, and the arrows mark the magnetic transition
temperatures.}
\end{figure}

\begin{figure}
\includegraphics[width=12cm]{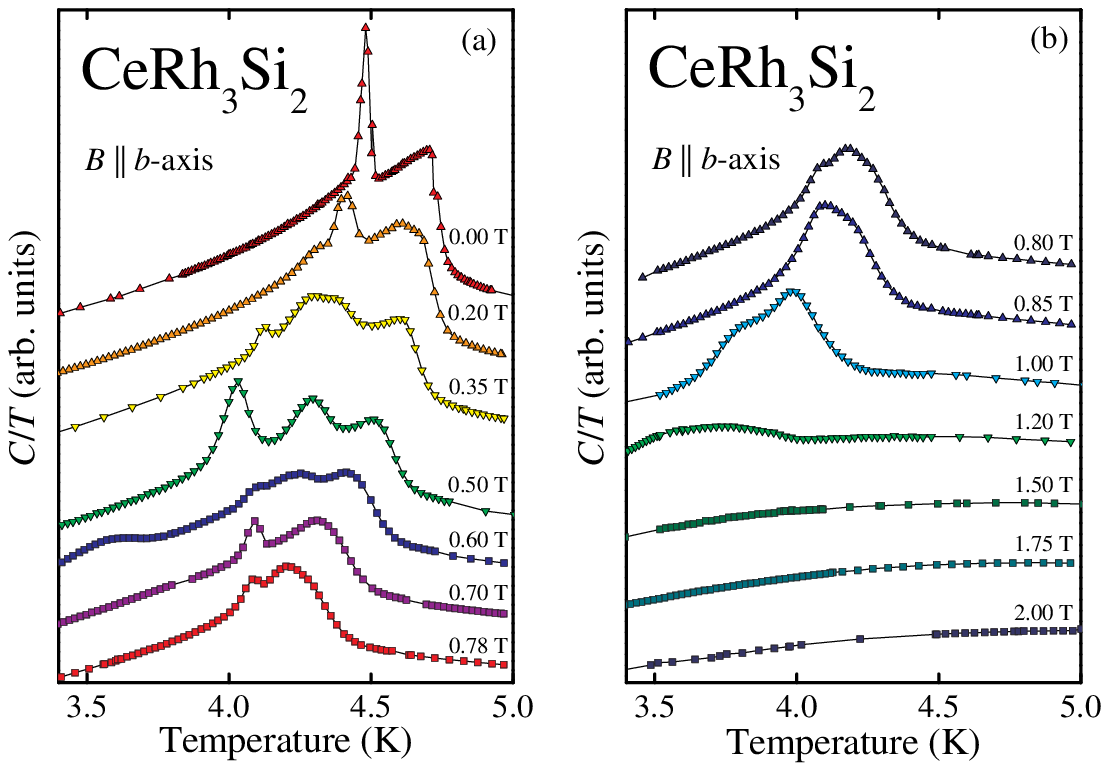}
\caption{\label{specific_heat_low_temperatures_figure} (Color online) Specific heat of
single-crystalline CeRh$_3$Si$_2$ measured at low temperatures in several different magnetic
fields $B$ applied along the $b$-axis. The curves are shifted up for the sake of clarity.}
\end{figure}

Figure~\ref{specific_heat_figure} displays the temperature dependencies of the specific heat of
CeRh$_3$Si$_2$ and its non-magnetic isostructural counterpart LaRh$_3$Si$_2$. At low temperatures a
distinct anomaly located at about 4.7~K confirms the antiferromagnetic ordering of the cerium
moments in CeRh$_3$Si$_2$. The overall shape of $C(T)$ obtained for the La-based compound is in
turn typical for non-magnetic metals. In particular, below about 7~K it is easily describable by
the formula:
\begin{equation}
\label{debye_low_temperatures_equation}
C (T) = \gamma T + \frac{12 \pi^4}{5} \frac{n R}{\Theta_{\rm D}^3} T^3,
\end{equation}
in which the first term is a conventional conduction-electron contribution to the specific heat
with $\gamma$ being the Sommerfeld coefficient, and the second term is a low-temperature phonon
contribution in a form of the Debye-$T^3$ law with the characteristic Debye temperature
$\Theta_{\rm D}$, the universal gas constant $R$, and $n$ being a number of atoms in the formula
unit [\onlinecite{gopal}]. Least-squares fitting of Eq.~(\ref{debye_low_temperatures_equation}) to
the experimental data (see the solid line in the inset to Fig.~\ref{specific_heat_figure}) yielded
the values of $\gamma$~=~3~mJ/(mol K$^2$) and $\Theta_{\rm D}$~=~428~K. At higher temperatures
$C(T)$ of LaRh$_3$Si$_2$ can be described by the equation:
\begin{equation}
\label{debye_einstein_gamma_equation}
C (T) = \gamma T + n_{\rm D} \times 9 R \left( \frac{T}{\Theta_{\rm D}} \right)^3
\int_0^{\Theta_{\rm D}/T} \frac{x^4 \; e^x}{(e^x - 1)^2} dx + n_{\rm E} \times 3 R \left(
\frac{\Theta_{\rm E}}{T} \right)^2 \frac{e^{\Theta_{\rm E}/T}}{\left( e^{\Theta_{\rm E}/T}
- 1 \right)^2},
\end{equation}
in which the second term is the full Debye expression for the phonon specific heat, and the third
term describes the phonon specific heat within the Einstein model of lattice vibrations (see e.g.
Refs.~[\onlinecite{gopal,tari}]) with $\Theta_{\rm E}$ being the characteristic Einstein
temperature. The coefficients $n_{\rm D}$ and $n_{\rm E}$ are numbers of atoms in the formula unit
($n_{\rm D} + n_{\rm E} \equiv n$), vibration of which were assumed to follow the Debye or Einstein
models, respectively. The solid line in Fig.~\ref{specific_heat_figure} presents the temperature
variation of the specific heat calculated employing Eq.~(\ref{debye_einstein_gamma_equation}) with
the parameters $\gamma$~=~3~mJ/(mol K$^2$) and $\Theta_{\rm D}$~=~428~K (as determined from the
low-temperature data), and $\Theta_{\rm E}$~=~125~K, $n_{\rm D}$~=~5 and $n_{\rm E}$~=~1. The two
latter values ($n_{\rm D}$ and $n_{\rm E}$) were chosen arbitrarily, assuming that the rare-earth
atoms (1/f.u.) in CeRh$_3$Si$_2$ and LaRh$_3$Si$_2$, which occupy the centers of the hexagonal
tubes (see Fig.~\ref{structure_figure} and Sec.~\ref{structure_section}), are weakly coupled with
their surroundings (5 at./f.u.) and exhibit some tendency to vibrate independently of the whole
lattice. The characteristic temperature of those vibrations (i.e. the Einstein temperature) was
estimated on the basis of simple analysis of $C(T)$  measured for the La-based compound: analytical
properties of the Debye and Einstein functions allow to relate the position $T_{\rm max}$ of the
low-temperature maximum in the $C(T)/T^3$ curve to the Einstein temperature as $\Theta_{\rm E}
\approx 5 \times T_{\rm max}$. In the case of LaRh$_3$Si$_2$ $T_{\rm max} \approx $~25~K, hence
$\Theta_{\rm E} \approx $~125~K (see the lower inset to Fig.~\ref{specific_heat_figure}).

In order to estimate a 4$f$-derived contribution $\Delta C(T)$ to the total heat capacity of
CeRh$_3$Si$_2$, the phonon specific heat of LaRh$_3$Si$_2$ was subtracted from $C(T)$ of the
Ce-based compound. As apparent from Fig.~\ref{specific_heat_magnetic_figure}(a), the so-obtained
$\Delta C (T)$ exhibits in the paramagnetic region a distinct Schottky-like anomaly due to CEF
effect. The thick solid line in Fig.~\ref{specific_heat_magnetic_figure}(a) represents the Schottky
specific heat calculated for the CEF model derived in Sec.~\ref{CEF_calc}. Apparently, the model
accounts reasonably well for the experimental data, except for high temperature region, where
significant quantitative discrepancy is seen. The shortcomings likely arise because of the
accumulation of experimental errors (see Fig.~\ref{specific_heat_figure}) directly influencing
reliability of the subtraction procedure.

Below about 50~K, the magnetic contribution to the specific heat of CeRh$_3$Si$_2$ increases again,
and at about 4.70(1)~K $\Delta C(T)$ takes a form of distinct $\lambda$-shaped anomaly, followed by
another pronounced peak, that occurs at about 4.48(1)~K. While the $\lambda$-shaped feature
corresponds to the second-order antiferromagnetic phase transition evidenced at $T_{\rm N}$ in the
magnetic susceptibility data, the second feature can be ascribed to the spin-rearrangement at
$T_{\rm t}$. The characteristic spike-like shape of that singularity is typical for first order
phase transitions. In turn, the magnetic contribution to the specific heat in the range 5--50~K can
be attributed to short-range interactions, which evolve finally into the long-range
antiferromagnetic ordering. The latter hypothesis is supported by the temperature dependence of the
magnetic entropy, defined as $S(T) = \int_0^T \Delta C(T)/T \, dT$. As can be inferred from
Fig.~\ref{specific_heat_magnetic_figure}(a) (right axis), the entropy at $T_{\rm N}$ is strongly
reduced in comparison to the entropy of the ground state doublet ($=R\ln{2}$) and amounts only to
about $0.44 \, R\ln{2}$. The value of $R\ln{2}$ is achieved at temperature as high as
$\approx$~30~K, suggesting that even well above $T_{\rm N}$ the system is not yet completely
disordered.

In the ordered region, the specific heat of CeRh$_3$Si$_2$ is dominated by an antiferromagnetic
magnon contribution. Assuming that the dispersion relation in the antiferromagnetic spin-wave
spectrum has a relativistic form $\omega = \sqrt{\Delta^2 + D^2k^2}$ (where $\omega$, $D$ and $k$
are the spin-wave frequency, stiffness, and wave number, respectively, and $\Delta$ is a spin-wave
gap) $\Delta C (T)$ takes the form of a sum of the electronic and magnon contributions
[\onlinecite{continentino}]:
\begin{equation}
\label{magnons_equation} \Delta C (T) = \gamma T + A \Delta^4 \sqrt{\frac{T}{\Delta}} e^{-\Delta /
T} \; \left[ 1 + \frac{39}{20} \left( \frac{\Delta}{T} \right) + \frac{51}{32} \left(
\frac{\Delta}{T} \right)^2 \right], \end{equation} where $A$ is a proportionality coefficient
related to $D$ as $A \propto 1/D^3$. Least-squares fitting procedure [see the solid line in
Fig.~\ref{specific_heat_magnetic_figure}(b)] yielded the values: $\gamma$~=~7~mJ/(mol~K$^2$),
$A$~=~3~mJ/(mol~K$^4$), and $\Delta$~=~16~K. The small value of the Sommerfeld coefficient, which
can be also directly deduced from Fig.~\ref{specific_heat_magnetic_figure}(b) as $\Delta C(T)/T
|_{\rm T \rightarrow 0}$, is typical for simple metals.

Figure~\ref{specific_heat_low_temperatures_figure} shows the results of the specific heat
measurements done in several different magnetic fields applied parallel to the $b$-axis. As in the
case of the magnetization measurements the latter experiments were also focused on the temperature
region close to $T_{\rm N}$ and $T_{\rm t}$. As seen, the two anomalies visible in zero magnetic
field at 4.7 and 4.5~K split in higher fields into four separate features in $C(T)/T$, which
independently shift to lower temperatures with rising $B$. However, all of the anomalies
systematically broaden and in $B \geqslant 0.7$~T one can distinguish only two of them. Finally, in
$B=$~1.2~T only one hump at about 3.7~K is visible, and in $B \geqslant$~1.5~T the specific heat
curve is featureless. The positions of the anomalies from
Fig.~\ref{specific_heat_low_temperatures_figure} served as complementary set of data used for
constructing the tentative phase diagram shown Sec.~\ref{summary}.

\subsection{CEF calculations}\label{CEF_calc}

\begin{table*}
\caption{\label{cef_parameters_table} CEF parameters, molecular field constant $\lambda$,
eigenenergies and eigenwavefunctions obtained from the fitting of the experimental magnetic
susceptibility data and the Schottky contribution to the specific heat.}
\begin{ruledtabular}
\begin{tabular}{l l}
Parameter & Value [K]\\ \hline
$B_{20}$ & 1322\\
$B_{22}$ & $-$1265\\
$B_{40}$ & 727\\
$B_{42}$ & 1327\\
$B_{44}$ & $-$2345\\
$B_{60}$ & $-$7478\\
$B_{62}$ & $-$2197\\
$B_{64}$ & 3060\\
$B_{66}$ & 22239\\
$\lambda$ [mol m$^{-3}$]& 501176\\
SQX & 1.86\%\\ \hline
Eigenvalues [K] & Eigenvectors \\ \hline
0 & $-\,0.801 \, |5/2,5/2\rangle + 0.428\,|5/2,1/2\rangle - 0.407 \, |5/2,-3/2\rangle$ \\
655 & $0.875 \, |5/2,3/2\rangle - 0.289\,|5/2,-5/2\rangle - 0.255\,|7/2,3/2\rangle +
$~\ldots \\
& \hspace{2cm}\dots~$+\,0.239\,|5/2,-1/2\rangle - 0.126\,|7/2,-1/2\rangle$\\
692 & $-\,0.783\,|5/2,1/2\rangle - 0.464\,|5/2,5/2\rangle - 0.334\,|7/2,1/2\rangle +
0.223\,|7/2,-7/2\rangle$\\
3091 & $-\,0.765\,|7/2,7/2\rangle - 0.511\,|7/2,-1/2\rangle + 0.313\,|7/2,3/2\rangle -
0.235\,|7/2,-5/2\rangle$\\
3478 & $-\,0.940\,|7/2,5/2\rangle + 0.246\,|7/2,1/2\rangle -
0.217\,|7/2,-3/2\rangle$\\
5467 & $-\,0.742\,|7/2,-1/2\rangle + 0.459\,|7/2,7/2\rangle - 0.232\,|7/2,3/2\rangle
+$~\ldots \\
& \hspace{2cm}\ldots~$-\,0.381\,|5/2,-1/2\rangle - 0.138\,|7/2,-5/2\rangle -
0.133\,|5/2,-5/2\rangle$\\
5882 & $-\,0.854\,|7/2,3/2\rangle - 0.383\,|7/2,7/2\rangle - 0.233\,|5/2,3/2\rangle
+$~\ldots\\
& \hspace{2cm}\ldots~$+\,0.202\,|5/2,-5/2\rangle + 0.170\,|7/2,-5/2\rangle$\\
\end{tabular}
\end{ruledtabular}
\end{table*}

The temperature dependence of the magnetic susceptibility (Sec.~\ref{magnetic_properties_section})
and the Schottky contribution to the heat capacity (Sec.~\ref{specific_heat_section}) of
CeRh$_3$Si$_2$ suggest predominantly localized character of the $f$-electrons and the importance of
definite sequence of the CEF levels. To capture the main features of the 4$f^1$ state split in the
CEF potential, we apply here the standard static description based on the renormalized Hamiltonian
projected on the space of the pure $f$-electron states. The approach is known to be quite
satisfactory for insulating systems, strongly supported by both qualitative and quantitative
theoretical studies [\onlinecite{mulak,05gajek}].

In the Wybourne notation [\onlinecite{wybourne}], the one-electron CEF potential of the C$_{2v}$
point  symmetry appropriate to the case of CeRh$_3$Si$_2$ reads :
\begin{equation}
\label{cef_hamiltonian}
H_{\rm CEF} = B_{20} \hat{C}_0^{(2)} + B_{22} \hat{C}_2^{(2)} + B_{40} \hat{C}_0^{(4)} + B_{42}
\hat{C}_2^{(4)} + B_{44} \hat{C}_4^{(4)} + B_{60} \hat{C}_0^{(6)} + B_{62} \hat{C}_2^{(6)} +
B_{64} \hat{C}_4^{(6)} + B_{66} \hat{C}_6^{(6)}.
\end{equation}
where $B_{kq}$ are the CEF parameters, which are to be adjusted to the experimental data and
$\hat{C}_q^{(k)}$ are the normalized spherical harmonics. Besides, we employ also a simplified,
angular-overlap-model (AOM) form of the CEF potential defined through matrix elements of the
effective CEF interaction (see for instance Ref.~\onlinecite{05gajek}):
\begin{equation}
\left\langle m\left|H_{\rm CEF}\right| m^{\prime}\right\rangle
=\sum_{\tau t}\sum_{\mu}D_{\mu
m}^{(3)\ast}(0,\Theta_{\tau t},\Phi_{\tau t})D_{\mu m^{\prime
}}^{(3)}(0,\Theta_{\tau t},\Phi_{\tau t})e_{\mu}^{\tau}. \label{e2v}
\end{equation}
In this equation, $D_{\mu m}^{(3)}(0,\Theta_{\tau t},\Phi_{\tau t})$ is the matrix element of the
irreducible representation $D^{(3)}$ of the rotation group, and $R_{\tau t}$, $\Theta_{\tau t}$,
$\Phi_{\tau t}$ are the angular coordinates of the nearest neighbor atom $\tau t$ expressed in the
coordination system set at the magnetic atom site. The index $\tau$ distinguishes various
chemically nonequivalent atoms and $t$ runs over nearest neighbors $\tau$ atoms. Thus, the whole
CEF effect is described by three intrinsic parameters $e_{\mu}^{\tau}$ with $\mu = 0(\sigma),
1(\pi), 2(\delta)$ for each chemically nonequivalent nearest neighbor atom $\tau$. In the present
case there are six silicon atoms and six rhodium atoms in the coordination sphere of the radius of
3.34~\AA. This coordination implies six non-equivalent AOM parameters, three for each nearest
neighbor atom. If we neglect $e_{\delta}$ and/or tie up $e_{\pi}$ with one third of $e_{\sigma}$
(following the rules observed for insulating systems [\onlinecite{mulak}]), four-parameters or
two-parameters versions of AOM can be applied.

The Hamiltonian $H$ includes also the spin-orbit interaction:
\begin{equation}
\label{so_hamiltonian}
H_{\rm SO} = \zeta_{4f} \mathbf{l}\cdot \mathbf{s}
\end{equation}
with the spin-orbit constant $\zeta_{4f}$ equal to 647.3 $cm^{-1}$ (931.4 K)
[\onlinecite{carnall}], and the Zeeman term:
\begin{equation}
\label{zeeman_hamiltonian}
H_{\rm M} = \mu_{\rm B} \left[ (l_x + 2 s_x) \frac{B_x}{B} + (l_y + 2 s_y) \frac{B_y}{B} + (l_z + 2 s_z)
\frac{B_z}{B} \right] B
\end{equation}
describing the system response to the external magnetic field $B$:
\begin{equation}
\label{hamiltonian}
H=H_{\rm CEF} + H_{\rm SO} + H_{\rm M}
\end{equation}

Following Refs.~\onlinecite{lueken} and \onlinecite{schilder}, the molar magnetization $M_{{\rm
mol}, \alpha}$ and the magnetic susceptibility $\chi_{\alpha}$ are calculated from derivative of
the free energy $F$ of the system:
\begin{equation}
\label{free_energy}
F= -\frac{N}{\beta} \sum_{\nu} \exp{\left( -\beta E_{\nu} \right)}; \beta = 1/(k_{\rm B}T)
\end{equation}
with respect to the magnetic field direction $\alpha$, {\em i.e.}:
\begin{equation}
\label{M_mol_a}
M_{{\rm mol}, \alpha} = -\frac{1}{V} \frac{\partial F}{\partial B} = N_{\rm A} \frac{\sum_{\nu}
\mu_{\nu, \alpha} \exp{\left( -\beta E_{\nu, \alpha} \right)}}{\sum_{\nu} \exp{\left( -\beta
E_{\nu, \alpha} \right)}}
\end{equation}
where
\begin{equation}
\label{mu_nu_a}
\mu_{\nu, \alpha} = -\frac{\partial E_{\nu, \alpha}}{\partial B}
\end{equation}
and $E_{\nu, \alpha}$ are the eigenenergies of the Hamiltonian given by Eq.~\ref{hamiltonian}. The
molar susceptibility is defined as:
\begin{equation}
\label{chi_mol_a}
\chi_{\alpha} = \mu_0 \frac{M_{{\rm mol}, \alpha}}{B}.
\end{equation}

The CEF parameters in Eq.~\ref{cef_hamiltonian} have been fitted to restore the temperature
dependences of the magnetic susceptibility measured along the $a$-, $b$- and $c$- axes and
simultaneously account for the Schottky contribution to the heat capacity (see section
\ref{specific_heat_section}). For this purpose the Levenberg-Marquardt method [\onlinecite{press}]
was applied in the version implemented in the computer programs by Helmut Schilder and Heiko Lueken
[\onlinecite{condon, lueken, schilder, schilderprivate}]. In order to scan the entire parametric
space, several initial CEF parameter sets have been probed. Among them one may distinguish three
special cases: (i) zero-values, (ii) the CEF parameters reported for the compound CeIr$_3$Si$_2$
[\onlinecite{shigetoh}] and (iii) the parameters estimated using AOM with the values:
$e_{\sigma}^{Si}$~=~$-$300,~300,~600~cm$^{-1}$ ($-$432, 432, 864 K), $e_{\pi}^{Si}$~=~$\pm
\frac{1}{3} e_{\sigma}^{Si}$, $e_{\delta}^{Si}$~=~$-$100, 0, 100~cm$^{-1}$ ($-$144, 0, 144 K) and
$e_{\sigma}^{Rh}$~=~$-$200, 200~cm$^{-1}$ ($-$288, 288 K), $e_{\pi}^{Rh}$~=~$\pm \frac{1}{2}
e_{\sigma}^{Rh}$, $e_{\delta}^{Rh}$~=~$-$50, 0, 50~cm$^{-1}$ ($-$72, 0, 72 K). Besides, several
independent initial randomly generated CEF parameters sets (over 230) have been probed as well.

The fitting procedure was same for each of the initial CEF parameters set. First, we fit the two
second-order parameters, whereas the remaining ones were kept at their initial values. Then, the
parameters of second and sixth rank were fixed at their values determined in the previous step,
while the fourth-rank parameters were fit. Subsequently, all the second-rank and the fourth-rank
parameters were varied simultaneously. In the following we call this initial step the base phase,
in which the general shapes of the experimental magnetic susceptibility curves have been reproduced
fairly well, yet some details of the measured curves could not be accounted for.

In the second, refining, phase we included the sixth-rank CEF parameters. Their influence on the
thermal properties is expected to be of less significance, as they only determine the CEF admixture
of the $7/2$ states having energies above 2000~cm$^{-1}$ (2878~K) to the low-energy $5/2$ states
predominantly governing the thermodynamic properties of the compound studied in moderate
temperature ranges. This admixture modifies rather slightly both the CEF splitting of the $5/2$
state and the matrix elements of the angular momentum operator. Nevertheless, the mixing may become
more pronounced if the CEF effect is strong enough. Indeed, some preliminary simulations showed
that in the case of CeRh$_3$Si$_2$ the admixture of the $j=7/2$ states may reach as much as 10\%
for the lowest states. Hence, we decided to include the sixth-rank states in the refining phase,
and as a result the relative root mean square error SQX
[\onlinecite{condon},\onlinecite{schilder04},\onlinecite{siddall76}] has dropped from 2.92\% to
1.86\%.

The fitting procedure of the sixth-order parameters were analogous to that of the fourth order.
First, we varied only the sixth-order parameters, while keeping all the remaining parameters fixed
on the values determined in the base phase. In the final step all the CEF parameters were varied
simultaneously. In each step the molecular field constant $\lambda$ was also varied.

Most of the probed initial sets of CEF parameters led to stable solutions of three different types.
The corresponding fits yielded similar sequences of the CEF levels but distinctly different
parameters sets, except for the two second-rank parameters, which appeared to have fairly similar
values in each solution. The best fitting results are collected in Tab.~\ref{cef_parameters_table}.
As seen from this table, the CEF effect is generally large and leads to the total splitting of the
$5/2$ state of the order of 500 cm$^{-1}$ (720 K) and the overall $4f^1$ spitting exceeding 4000
cm$^{-1}$ (5755 K). Despite the pronounced differences between the parameters in the three distinct
sets (only the solution giving lowest SQX is shown) the sequence of levels and the model curves are
almost same, what is reflected in negligible differences in the root-square errors and the
energies. The scattering of some fourth- and especially sixth-rank CEF parameters in various
solutions confirms rather limited reliability of the latter parameters. Consequently, as the final
result we accepted the solution characterized by the lowest root-square error. The so-obtained
model curves are shown in Fig.~\ref{susceptibility_figure} together with the experimental ones.
Additionally, Fig.~\ref{specific_heat_magnetic_figure}(a) presents the calculated Schottky
contribution expected for the obtained CEF splitting scheme.

\subsection{\label{XPS experiment} X-ray photoemission spectroscopy}

Analysis of the Ce $3d$ XPS spectra is a proper method to investigate the character of the Ce $4f$
states in Ce-intermetallic compounds due to the strong Coulomb interaction between photoemission
holes in the $3d$ shell and electrons located near the Fermi level. $3d$ core ionization of
Ce-based compounds usually results in a spectrum of final states, whose configurations correspond
to the $f^0$, $f^1$, and $f^2$ states (see Refs. \onlinecite{Gun83} and \onlinecite{Fug83}).

\begin{figure}
\centering
\includegraphics[width=8.6cm]{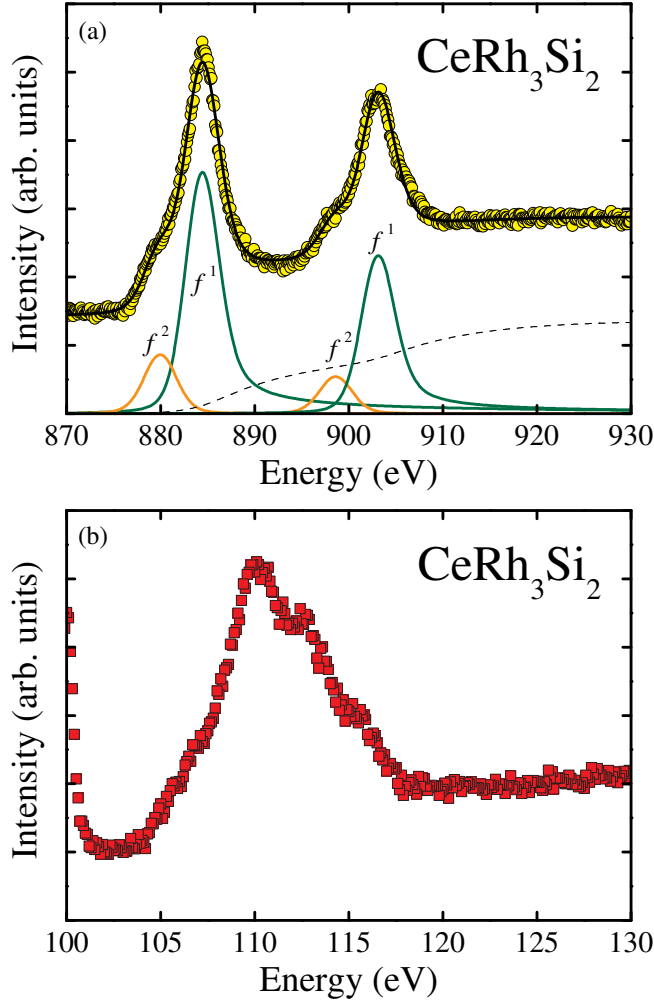}
\caption{\label{XPS} (color online) (a) Ce 3\textit{d} and (b) Ce 4\textit{d} XPS spectra
obtained for CeRh$_3$Si$_2$. The $f^{1}$ and $f^{2}$ components were separated on the basis of the
Doniach-\v Sunji\'c theory.}
\end{figure}

Figure \ref{XPS}(a) shows the Ce $3d$ XPS spectrum obtained for CeRh$_3$Si$_2$. Two final-state
contributions $f^1$ and $f^2$ are clearly observed, which exhibit spin-orbit splitting
$\Delta_{SO}\cong 18.6$ eV. In turn, the $f^0$ component is not evident in the spectrum, which
implies stable valence $3+$ of the cerium atoms. According to the Gunnarsson and Sch\"onhammer
theory [\onlinecite{Gun83}], the hybridization energy $\Delta_{fs}$, which describes the
hybridization part of the Anderson impurity Hamiltonian [\onlinecite{And61}] and is defined as $\pi
V^{2}\tilde{\rho}_{max}$ [where $\tilde{\rho}_{max}$ is the maximum in the density of states (DOS)
and $V$ is the hybridization matrix element], can be estimated from the ratio
$r=I(f^{2})/(I(f^{1})+I(f^{2}))$ of the $I(f^{2})$ and $I(f^{1})$ intensities [\onlinecite{Fug83}].
The particular contributions to the 3$d$ XPS spectrum can be derived by its analysis in terms of
the Doniach-\v Sunji\'c approach [\onlinecite{Don70}] (the experimental details and accuracy of the
method is also discussed in Ref. \onlinecite{Sle04}). This way, for CeRh$_3$Si$_2$ one estimates a
hybridization width $\Delta_{fs}$ of $\sim$56 meV.

More evidence for the stable valence of Ce$^{3+}$ ions in CeRh$_3$Si$_2$ comes from the Ce 4$d$ XPS
spectrum displayed in Fig. \ref{XPS}(b). Clearly, there is no additional peak at higher binding
energies ($E>118$ eV) that could be attributed to the Ce $3d^9f^0$ state
[\onlinecite{Sig83,Bae78,Fug83}].

\subsection{\label{band structure} Electronic band structure calculations}

The band structure calculations were performed using the Wien2k code within the LDA, LDA+U and
LSDA+OP schemes [the first two approaches with (LSDA and LSDA+U) and without spin polarization].
Non spin polarized densities of electronic states were used for calculations of photoemission
spectra, which in experiment were measured at room temperature, well above T$_N$=4.72 K. In the
case of spin polarized calculations the antiferromagnetic solution was not assumed in advance. The
twelve atoms forming unit cell that was used in our calculations were treated as different types of
atoms. The starting magnetic moments were chosen to have opposite sign. The system reached,
iteration by iteration, a selfconsistent solution that was very close to the antiferromagnetic
state. The resultant magnetization of both sublattices was almost balanced and was equal to about
2x10$^{-3}$ $\mu_B$ per f.u..

\begin{figure}
\centering
\includegraphics[width=8.6cm]{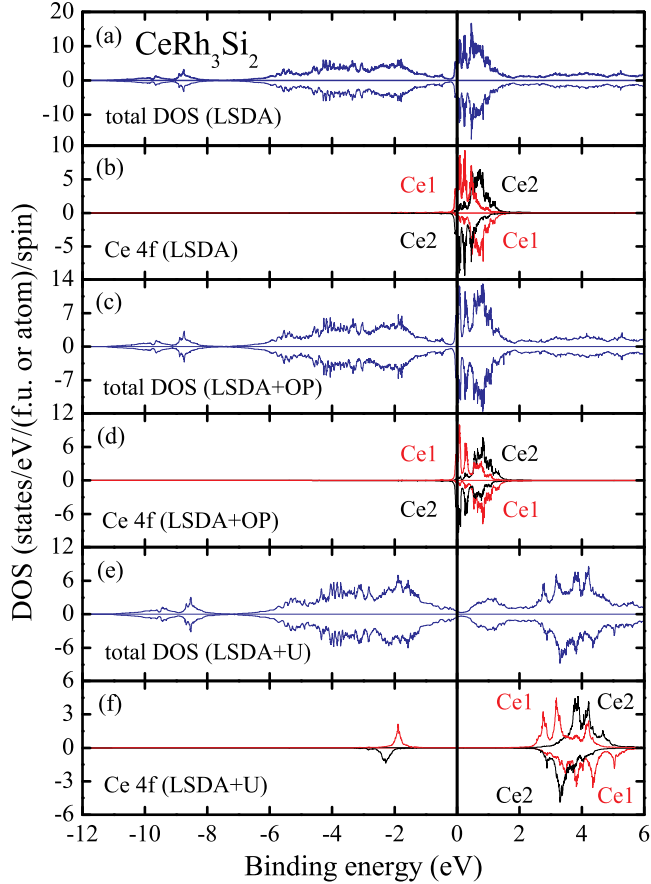}
\caption{\label{band_structure_figure}The total (per f.u.) and site projected (per Ce atom)
densities of electronic states (DOS) for antiferromagnetic
CeRh$_3$Si$_2$ calculated within the LSDA, LSDA+OP, and LSDA+U
schemes.}
\end{figure}

In the first step of our calculations, we employed the LSDA approach. The calculated densities of
states (DOS) for CeRh$_3$Si$_2$ system are presented in Fig.\ref{band_structure_figure}: (a) the
total DOS and (b) the contributions provided by 4f electrons located on cerium atoms forming two
magnetic sublattices. The valence band can be divided into two parts below the Fermi level (E$_F$),
here located at E=0. The first part located in the range of (-11.5;-7.5) eV is formed mainly by
electrons provided by silicon atoms Si(3s). The main part of the valence band between about -7.5 eV
and E$_F$ is created by Rh(4d) and Si(3p) electrons. The Ce(4f) electrons provide contribution
nearest the Fermi level, and they have the main contribution to DOS at E=E$_F$: 5.43 states/(eV
atom), 77.5\% of the total value 7.01 states/(eV f.u.).

Based on DOS at E$_F$ one can calculate the Sommerfeld coefficient $\gamma_0$, which in our case is
equal to about 16.5 mJ/(mol K$^2$). This value is much higher than the experimental one
$\gamma_{exp}$=3 mJ/(mol K$^2$), reported earlier in Sec. IIID. This discrepancy is probably caused
by inadequate description of the strongly correlated 4f electrons by the LSDA formalism.

Hybridization changes initial configuration 6s$^{2}$5d$^{1}$4f$^{1}$ to the final one
6s$^{1.99}$5d$^{0.57}$4f$^{1.02}$. Total number of electrons is different from those for pure
cerium systems because of charge transfer from/to rhodium and silicon atoms as well as the charge
accumulated in the interstitial region between atomic spheres.

The values of site projected moments on cerium atoms are equal to +/-0.08 $\mu_B$ per Ce atom. The
resultant value is not spectacular because of opposite alignment of spin and orbital moments. Until
now the experimental values of local moments were not known.

The shape of DOS obtained by LSDA+OP approach (Fig.~\ref{band_structure_figure} (c) and (d)) is
similar to that reported earlier but usually in this approach the magnetic moments increase, in our
case to +/-0.5 $\mu_B$ per Ce atom. Unfortunately the DOS at E$_F$ is still too high: 10.65
states/(eV f.u.) with contribution of 4f electrons of about 8.52 states/(eV Ce atom). The
Sommerfeld coefficient is equal to 25.1 mJ/(mol K$^2$).

On the two lowest panels of Fig.~\ref{band_structure_figure} we present results of LSDA+U approach
with U equal to 6 eV. The additional U term to the LSDA approach shifts the 4f DOS (in LSDA located
well below E$_F$) towards higher binding energies (now located around -2 eV). The unoccupied part
of 4f DOS is moved to the energies about 4 eV above the Fermi level. In this way the contribution
of Ce(4f) electrons to the DOS at E$E_F$ is practically negligible, and the Fermi level is located
in a pseudo gap of the DOS. The value of DOS(E$_F$)=1.01 states/(eV f.u.) gives the Sommerfeld
coefficient equal to 2.39 mJ/(mol K$^2$). The latter value  is slightly smaller than the
experimental Sommerfeld coefficient as it should be for theory. The magnetic moments are equal to
+/-0.09 $\mu_B$ per Ce atom.

\begin{figure}
\centering
\includegraphics[width=8.6cm]{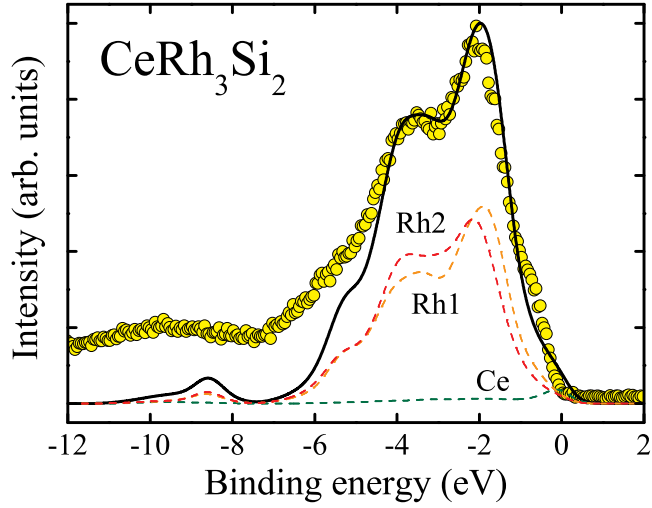}
\caption{\label{valence_band_figure} The measured X-ray photoelectron spectrum for the valence
band compared with the calculated ones within the LDA approach for CeR$_3$hSi$_2$ system.}
\end{figure}

Figure~\ref{valence_band_figure} presents experimental and calculated X-ray photoelectron spectra
based on non spin polarized DOS plots and calculated cross sections collected in
Ref.~\onlinecite{as09}. The best fitting to the experimental photoemission spectrum was obtained
for the LDA calculations. The consistency between the calculated and experimental spectra is quite
good. The spectrum is dominated by the Rh(4d) electrons. The Ce(4f) electrons provide contributions
near the Fermi level. The peak located at about 8-9 eV below E$_F$ is formed mainly by the Si(3s)
electrons (not shown here).

\section{\label{summary} Summary}

\begin{figure}
\includegraphics[width=8.6cm]{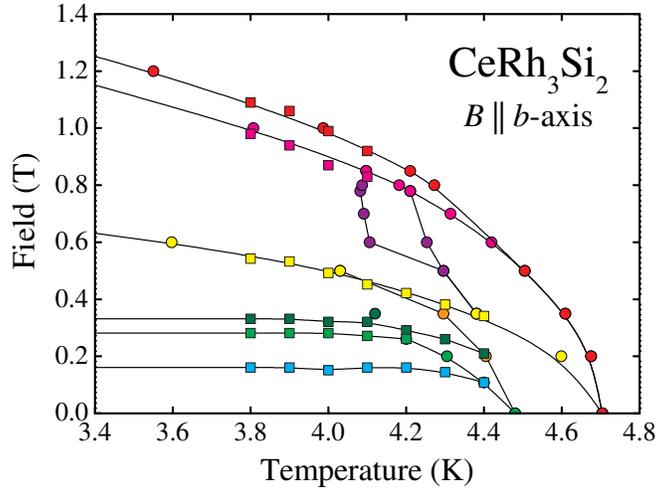}
\caption{\label{phase_diagram_figure} (Color online) Tentative $B\!-\!T$ phase diagram for
single-crystalline CeRh$_3$Si$_2$ based on anomalies observed in temperature dependencies of
magnetic susceptibility (squares; Fig.~\ref{susceptibility_low_temperatures_figure}),
field variations of magnetization (squares;
Fig.~\ref{magnetization_low_temperatures_figure}) and the temperature dependencies of the
specific heat (circles; Fig.~\ref{specific_heat_low_temperatures_figure}) for $B\parallel
b$-axis. Solid curves are hypothetical phase-boundary lines. Colors serves as a guide for the eye.}
\end{figure}

CeRh$_3$Si$_2$ was found to crystallize in the orthorhombic ErRh$_3$Si$_2$-type structure with
honey-comb-like Rh-Si cages surrounding the Ce ions, which occupy just one inequivalent
crystallographic position in the unit cell. The $4f$ electrons are well localized and experience
exceptionally strong (as for cerium intermetallics) crystalline--electric--field effect with
significant admixture of two $j$-multiplets (i.e. 5/2 and 7/2). In the whole temperature range
covered CeRh$_3$Si$_2$ exhibits very large easy $ab$ plane anisotropy in its magnetic and
electrical transport behavior. At low temperatures the compound undergoes two subsequent
antiferromagnetic phase transitions. The experiments performed in external magnetic fields $B$
applied parallel to the easy $b$-axis revealed very complex $B-T$ phase diagram of CeRh$_3$Si$_2$
(Fig.~\ref{phase_diagram_figure}), which needs to be verified by neutron-scattering studies. The
band structure calculations confirmed the antiferromagnetic ordering of the cerium magnetic
moments. The calculated values of these moments are equal to 0.08, 0.09, and 0.5 $\mu_B$ per Ce
atom for the LSDA, LSDA+U, and LSDA+OP approach, respectively. Only LSDA+U calculations yielded
correct relation between theoretical [2.39 mJ/(mol K$^2$)] and measured [3 mJ/(mol K$^2$)] values
of the Sommerfeld coefficients being less than one. The non-spin-polarized DOS plots reproduced
well the experimental X-ray photoelectron spectrum of the valence band.

It is worth noting that properties similar to those of CeRh$_3$Si$_2$ (i.e. "devilish"
magnetization, complex magnetic phase diagram, strong CEF effect) were reported for a number of
magnetically ordered rare-earth intermetallics. Among them one can mention highly anisotropic
CeIr$_3$Si$_2$ [\onlinecite{shigetoh}], TbNi$_2$Ge$_2$~[\onlinecite{budko, andre}],
TbNi$_2$Si$_2$~[\onlinecite{shigeoka}] and CeSb~[\onlinecite{rossat-mignod1, rossat-mignod2}]. All
these compounds exhibit several successive field-induced metamagnetic phase transitions, which
manifest themselves as a series of sudden jumps between well-defined plateaus of field-dependent
magnetization. Detailed bulk direction-dependent physical properties measurements as well as
neutron diffraction and x-ray resonant exchange scattering studies have revealed that in all of
these compounds one deals with non-trivial field-induced spin rearrangements. In most cases the
metamagnetic processes are relatively well understood and described phenomenologically, yet
microscopic models were proposed only for a limited number of the systems (for survey see
Ref.~\onlinecite{gignoux} and references therein). In most cases, the magnetic behavior was
interpreted in terms of the Ising model with next-nearest neighbour interactions (Axial
Next-Nearest Neighbor Ising model, ANNNI), which (in an ideal case) can produce an infinite number
of commensurate phases occurring with an infinite number of steps in field-dependent magnetization,
called commonly a "devil's staircase" (for review see \emph{e.g.} Ref.~\onlinecite{bak}). Other
theoretical approaches were proposed for TmAgGe [\onlinecite{morosan}] and HoNi$_2$B$_2$C
[\onlinecite{canfield1}]. The former compound has been found to exhibit some features of frustrated
system with distorted antiferromagnetic Kagome-like layered structure. The sequence of
field-induced metamagnetic transitions observed in the compound TmAgGe was successfully reproduced
using a hamiltonian that describes only (i) relatively weak antiferromagnetic next-nearest-neighbor
and ferromagnetic nearest-neighbor exchange adopted to the geometry of the system, and (ii) strong
crystal electric field effect~[\onlinecite{goddard}]. In turn, HoNi$_2$B$_2$C is a magnetically
ordered superconductor, which shows at low-temperatures a transition from commensurate
antiferromagnetic to incommensurate $c$-axis complex spiral state with some $a$-axis incommensurate
modulation~(see Refs.~\onlinecite{cho,rathnayaka,canfield2} and references therein). Here again the
crystalline electric field was taken into account to explain the complex magnetic phase
diagram~[\onlinecite{amici}], yet the generalized ANNNI model (a kind of so-called clock model) was
also successfully applied~[\onlinecite{kalatsky}].

The microscopic origin of the complex magnetic behavior of CeRh$_3$Si$_2$ cannot be defined at this
stage of our study. The most relevant factor governing its strongly anisotropic properties in the
ordered and paramagnetic states is clearly unusually strong crystalline electric field potential
that yield the total splitting of the 5/2 state of about 720 K and the overall 4$f^!$ splitting as
large as about 5760 K. The estimated magnitude of the CEF effect in CeRh$_3$Si$_2$ is perhaps the
largest ever reported for Ce-based intermetallics. The other cerium compounds known for their giant
CEF interactions are CeRh$_3$B$_2$ [\onlinecite{galatanu,givord}] and CeIr$_3$Si$_2$
[\onlinecite{shigetoh}], where the total CEF splitting is about 2000~K and 1600~K, respectively.
Remarkably, the latter compound is isostructural to CeRh$_3$Si$_2$, while the crystal structure of
CeRh$_3$B$_2$ is closely related to that of CeRh$_3$Si$_2$. Further studies, including inelastic
neutron scattering experiments and ANNNI calculations, are needed to shed more light on the unique
features (i.e. unusually strong CEF and complex magnetic behavior) of the latter compound.

\begin{acknowledgments}
We are grateful to Dr. Helmut Schilder for developing his computer program towards simultaneous
fitting of the heat capacity and for the engagement and support. This work was supported by the
Ministry of Science and Higher Education within grants no. N202 116 32/3270 and N202 1349 33. Part
of the research was performed in the frame of the National Network "Strongly correlated materials:
preparation, fundamental research and applications".
\end{acknowledgments}


\end{document}